\documentclass[aps,prb,twocolumn,letterpaper,superscriptaddress,showpacs]{revtex4}
\usepackage{CJK}
\usepackage{keyval}%
\usepackage{graphicx}
\usepackage{dcolumn}
\usepackage{bm}
\usepackage{color}
\usepackage{amsmath}
\usepackage{soul}

\setkeys{Gin}{width=0.8\columnwidth,clip=true}
\newcommand{\ignore}[1]{}

\begin{document}
\begin{CJK*}{UTF8}{bsmi}
\title{Band structure of silicene on the zirconium diboride (0001) thin film surface - convergence of experiment and calculations in the one-Si-atom Brillouin zone} 

\author{Chi-Cheng Lee (%
李啟正
)}
\affiliation{School of Materials Science, Japan Advanced Institute of Science and Technology (JAIST),
1-1 Asahidai, Nomi, Ishikawa 923-1292, Japan}%
\author{Antoine Fleurence}
\affiliation{School of Materials Science, Japan Advanced Institute of Science and Technology (JAIST),
1-1 Asahidai, Nomi, Ishikawa 923-1292, Japan}%
\author{Yukiko Yamada-Takamura
}
\affiliation{School of Materials Science, Japan Advanced Institute of Science and Technology (JAIST),
1-1 Asahidai, Nomi, Ishikawa 923-1292, Japan}%
\author{Taisuke Ozaki
}
\affiliation{School of Materials Science, Japan Advanced Institute of Science and Technology (JAIST),
1-1 Asahidai, Nomi, Ishikawa 923-1292, Japan}%
\affiliation{Institute for Solid State Physics, The University of Tokyo, Kashiwa 277-8581, Japan}%
\author{Rainer Friedlein
}
\affiliation{School of Materials Science, Japan Advanced Institute of Science and Technology (JAIST),
1-1 Asahidai, Nomi, Ishikawa 923-1292, Japan}%

\date{\today}

\begin{abstract}
So far, it represents a challenging task to reproduce angle-resolved photoelectron (ARPES) spectra of epitaxial silicene by first-principles calculations. Here, we report on the resolution of the previously controversial issue related to the structural configuration of silicene on the ZrB$_2$(0001) surface and its band structure. In particular, by representing the band structure in a large Brillouin zone associated with a single Si atom, it is found that the imaginary part of the one-particle Green's function follows the spectral weight observed in ARPES spectra. By additionally varying the in-plane lattice constant, the results of density functional theory calculations and ARPES data obtained in a wide energy range converge into the ``planar-like'' phase and provide the orbital character of electronic states in the vicinity of the Fermi level. It is anticipated that the choice of a smaller commensurate unit cell for the representation of the electronic structure will be useful for the study of epitaxial two-dimensional materials on various substrates in general.
\end{abstract}


\maketitle
\end{CJK*}

\section{Introduction}

Silicene, the Si version of the two-dimensional carbon allotrope graphene, promises interesting electronic properties that derive from its graphene-like electronic structure, a large spin-orbit coupling and the out-of-plane buckling of atoms belonging to the two distinguishable sub-lattices.\cite{Ref1,Ref2,Ref3,Ref4,Ref5} As such, Dirac fermions at the Fermi level ($E_F$),\cite{Ref3,Ref4} a large quantum spin Hall effect,\cite{Ref1,Ref2} a topological quantum phase transition\cite{Ref6} and perfect spin filtering \cite{Ref7} are predicted to occur in its low-buckled, free-standing form. Importantly, as compared to graphene, under varying external conditions, silicene is structurally more flexible and can occur with a variety of lattice constants, atomistic structures and with a varying $sp^2/sp^3$ ratio.\cite{Ref5,Ref8} This flexibility may allow tuning of the electronic properties\cite{Ref8,Ref9} to be adapted in useful applications.

While free-standing silicene is yet hypothetical, two-dimensional Si honeycomb lattices have recently been prepared on several metallic substrates such as Ag(111),\cite{Ref10,Ref11,Ref12} ZrB$_2$(0001)\cite{Ref8} and Ir(111)\cite{Ref13} which opens the way for their systematic and comprehensive experimental characterization. The structural and electronic properties of epitaxial silicene phases are expected to be significantly altered from those of free-standing silicene, and are determined by the strength of hybridization with in particular the $d$ electronic states of the substrates and the resulting types of buckling.\cite{Ref5,Ref14} Since this is a complex matter, not surprisingly, the interpretation of in particular observed surface reconstructions\cite{Ref10,Ref11,Ref12,Ref15,Ref16,Ref17} and of the spectral features visible in angle-resolved photoelectron (ARPES) spectra\cite{Ref8,Ref12,Ref18,Ref19} have been and is still a matter of controversy.

In contrast to multiple epitaxial silicene phases observed by scanning tunneling microscopy (STM) on the Ag(111) surface, so far, only a single phase has been reported to be formed on the ZrB$_2$(0001) surface.\cite{Ref8} In this phase, ($\sqrt{3}\times\sqrt{3}$)-reconstructed silicene is in an epitaxial relationship to the ($2\times2$) ZrB$_2$(0001) unit cell. While evidence for atomic-scale buckling comes from photoelectron diffraction data,\cite{Ref8} ARPES spectra show two upwards curved spectral features that approach $E_F$ by up to 250 meV at the $K_{Si}$($1\times1$) point of (unreconstructed) silicene and, owing to backfolding, are mirrored with weak intensity at the $\overline{\Gamma}$ point in the $1^{st}$ Brillouin zone (BZ).\cite{Ref8,Ref20}

\begin{figure}[tbp]
\includegraphics[width=0.95\columnwidth,clip=true,angle=0]{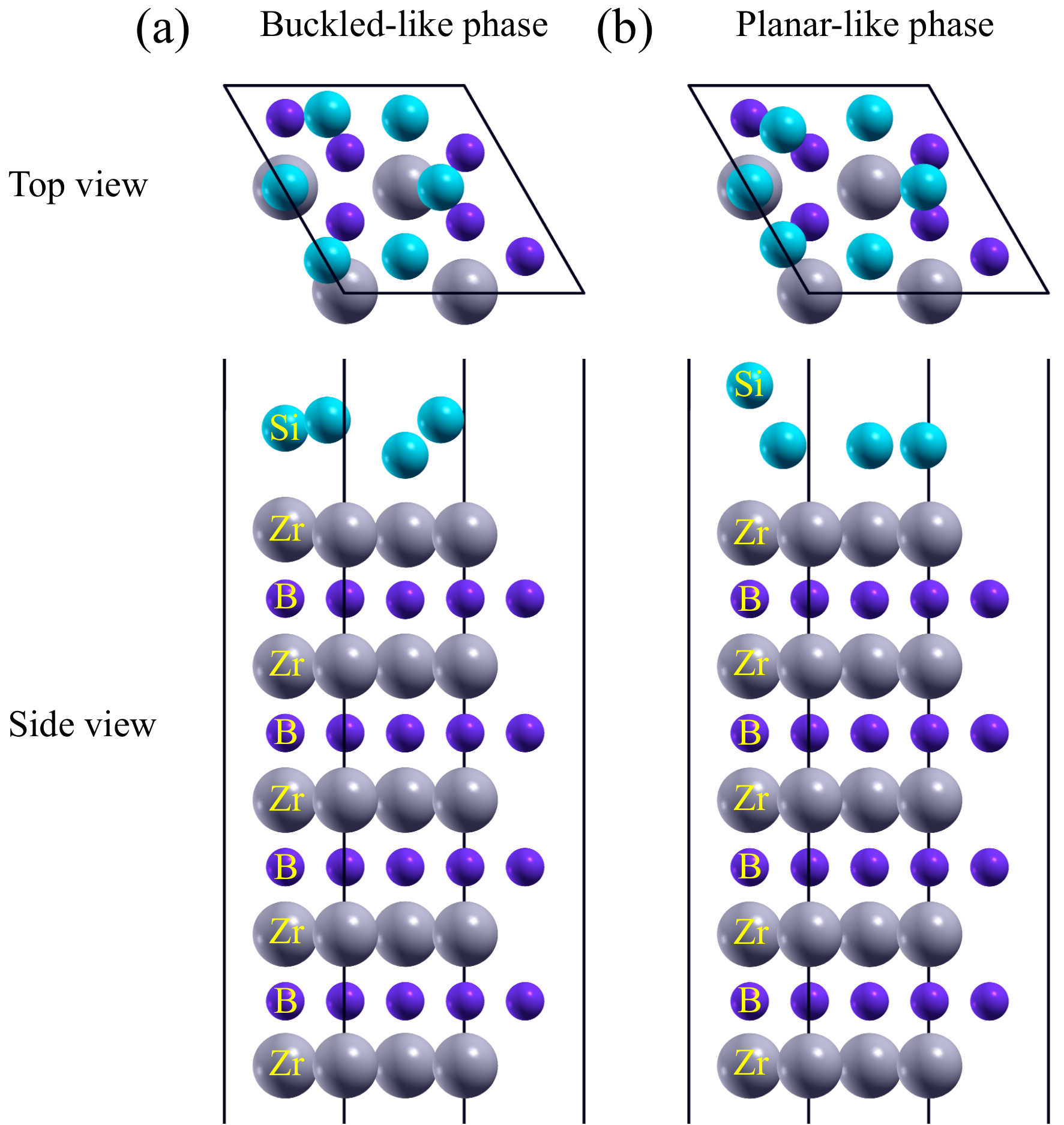}
\caption{\label{fig:fig1}
Structures of (a) buckled-like and (b) planar-like phases of ($\sqrt{3}\times\sqrt{3}$)-reconstructed silicene on the Zr-terminating ZrB$_2$(0001) surface.
}
\end{figure}

Density functional theory (DFT) calculations find two possible structures,\cite{Ref5} shown in Fig.~\ref{fig:fig1}, which so far only agree partially with the experimental data. In particular, while within DFT, a ``planar-like'' structure (Fig.~\ref{fig:fig1} (b)) was reported to be the ground state of silicene on ZrB$_2$(0001),\cite{Ref5} scanning tunneling spectroscopy (STS) spectra and photoelectron diffraction data could also be explained by the calculated electronic properties of the metastable, so called ``buckled-like'' phase (Fig.~\ref{fig:fig1} (a)).\cite{Ref8,Ref21} Even more, so far, the band structure of both phases shows only partial agreement with ARPES spectra.\cite{Ref5} This has preliminary been attributed to the presence of long-range interactions leading to the spontaneous formation of stress domains within the two-dimensional layer of Si atoms.\cite{Ref8}

In this context, another important issue that needs consideration is the sensitivity of the ARPES cross-section to the selected Brillouin zone.\cite{Ref22,Ref23} As such, the strong change of the graphene $\pi$ band intensity upon crossing the zone boundary\cite{Ref24,Ref25} has initially been considered to derive from the dependence of the photoelectron emission matrix elements on the sign of the initial state electron momentum\cite{Ref22} and has been discussed to be related to quantum mechanical interference between atoms in the two different sub-lattices.\cite{Ref23} In the case of epitaxial silicene on ZrB$_2$(0001), Si atoms are not in two sub-lattices but occupy three distinguishable atomic sites.\cite{Ref8,Ref26} However, similar interference effects may occur.

In the present calculations, we have introduced a conceptual unit cell that contains only a \textit{single} Si atom to allow interference between the distinguishable Si atoms defined in the ($\sqrt{3}\times\sqrt{3}$) unit cell via the phase factors in terms of unfolding the electronic band structures.\cite{Ref34,Ref35} By additionally varying slightly the in-plane lattice constant, we find good agreement between ARPES spectra obtained in a wide energy range and the calculated band structure of the planar-like phase, even without explicitly considering stress domains. The agreement allows for an analysis of the orbital character of the electronic states and reveals that the upwards curved bands which, according to ARPES data,\cite{Ref8} approach $E_F$ at $K_{Si}$($1\times1$) point are formed by hybridization of Si $sp^2$ and $p_z$ orbitals and contain contributions of Zr $d$ orbitals as well.

\section{Computational and Experimental}

The DFT calculations within a generalized gradient approximation (GGA)\cite{Ref27,Ref28} were performed using the OpenMX code based on norm-conserving pseudopotentials generated with multi reference energies\cite{Ref29} and optimized pseudoatomic basis functions.\cite{Ref30,Ref31} For each Zr atom, three, two, and two optimized radial functions were allocated for the $s$-, $p$-, and $d$-orbitals, respectively, as denoted by $s3p2d2$. For both Si and B atoms, $s2p2d1$ basis functions were adopted. The cut-off radius of 7 Bohr was chosen for all the basis functions. The regular mesh of 220 Ry in real space was used for the numerical integrations and for the solution of the Poisson equation. An ($8\times8\times1$) mesh of $k$ points was used to study an isolated slab consisting of five Zr, four B, and one silicene layers implemented by the effective screening medium method.\cite{Ref32} The force on each atom was relaxed to be less than 0.0001 Hartree/Bohr. The XCrySDen software was used to generate the figures.\cite{xcrysden}

ARPES spectra were obtained using the SES-200 hemispherical analyzer at the end-station of the undulator beam line 13B at the Photon Factory synchrotron radiation facility, located at the High Energy Accelerator Research Organization (KEK), Tsukuba, Japan, using the photon energy of $h\nu$ = 43 eV. The total energy resolution was better than 35 meV as determined from the broadening of $E_F$. At this end-station, the electric field vector of the light was at the fixed angle of 25$^{\circ}$ with respect to the photoelectron analyzer.

Oxide-free silicene samples were prepared as described previously.\cite{Ref8,Ref33} Note that under optimal conditions,\cite{Ref33} this procedure highly reproducibly leads to samples with more than 95\% of the surface covered with single-crystalline-like silicene.\cite{Hubault} In short, after annealing at about 780$^{\circ}$C under ultra-high vacuum conditions, silicene forms \textit{in situ} and spontaneously through surface segregation on metallic single-crystalline zirconium diboride (0001) thin films on Si(111) wafers. The ($2\times2$) reconstruction of the annealed ZrB$_2$(0001) surface and the characteristic chemical shifts of the Si 2$p$ core level components\cite{Ref8,Ref26} were verified by low-energy electron diffraction and surface-sensitive core level photoelectron spectroscopy using $h\nu$ = 130 eV. During the measurements, samples were held at room temperature.

\section{Results and Discussion}

\subsection{The one-atom unit cell}

As a consequence of the buckling and in-plane lattice distortions, the translational symmetry of ($1\times1$) silicene on the ZrB$_2$(0001) surface is broken such that silicene 
adapts the ($\sqrt{3}\times\sqrt{3}$) unit cell that is commensurate with the ($2\times2$)-reconstructed unit cell of the zirconium diboride surface. However, the degree of translational symmetry breaking might not be strong enough to allow ARPES experiments to observe the spectral weight calculated in this crystallographic unit cell. Therefore, in order to reflect the strength of translational symmetry breaking, the spectral weight as derived from the imaginary part of the one-particle Kohn-Sham Green's function may be unfolded to a larger Brillouin zone.\cite{Ref34,Ref35} For example, with the choice of the one-C-atom Brillouin zone of graphene, the photoelectron emission matrix element is basically the atomic form factor to be consistent with the alternative picture of destructive quantum mechanical interference between two C atoms.\cite{Ref23,Ref36} For silicene on the ZrB$_2$(0001) surface, we can choose to represent silicene in the one-Si-atom unit cell while keeping the Si atoms at their respective positions typical either of the ``buckled-like'' or ``planar-like'' phases. This of course leaves the band structure itself unchanged, and affects only the distribution of the spectral weight in the reciprocal space.

\begin{figure}[tbp]
\includegraphics[width=1.00\columnwidth,clip=true,angle=0]{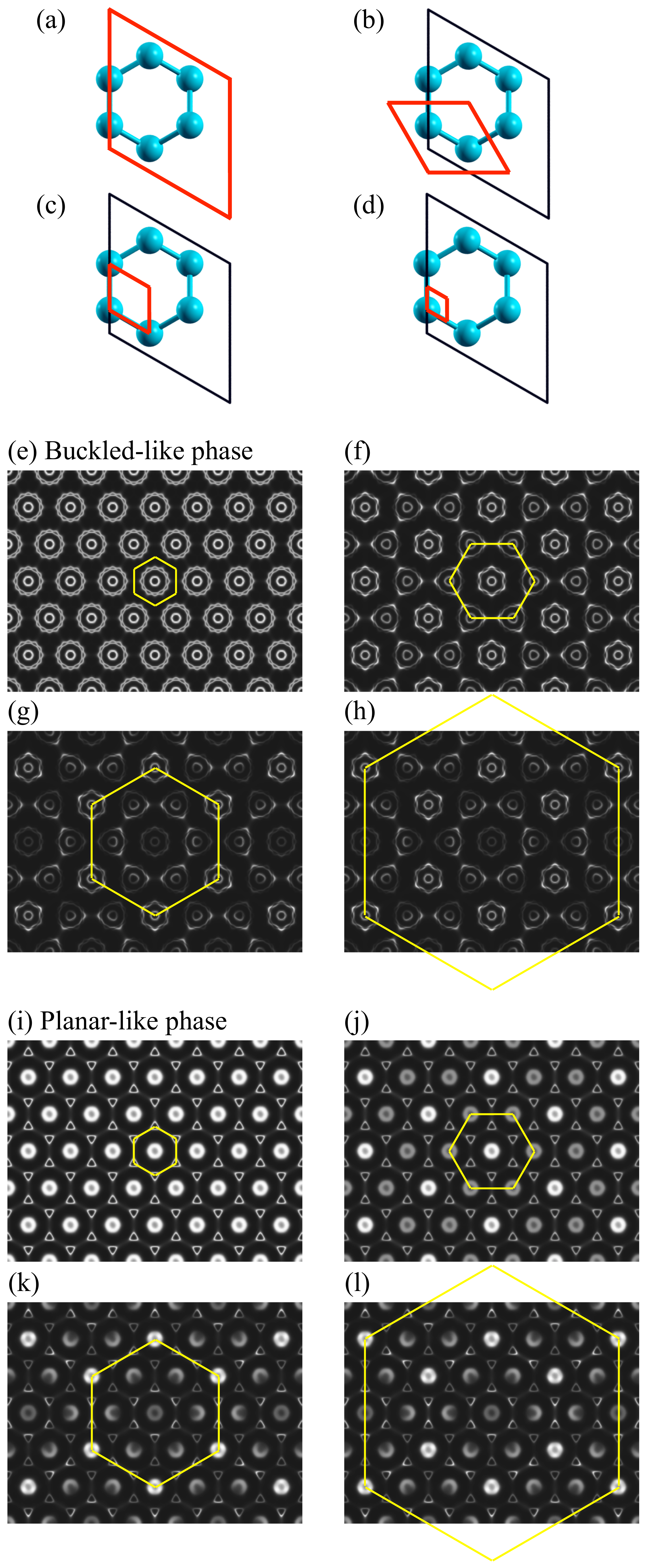}
\caption{\label{fig:fig2}
Spectral weight on the constant-energy plane of 1.0 eV below $E_F$ for (e-h) the buckled-like phase and (i-l) the planar-like phase represented in different Brillouin zones corresponding to (a-d) the adopted unit cells, respectively. The silicene structures are thought to be peeled off from those presented in Fig.~\ref{fig:fig1}.
}
\end{figure}

To illustrate the change of spectral weight associated with different conceptual unit cells, we first focus on isolated silicene sheets as thought to be peeled off from the optimized structures on the ZrB$_2$ substrate. We emphasize that the conceptual unit cells could be smaller than the primitive ($\sqrt{3}\times\sqrt{3}$) unit cell and the detailed description of conceptual unit cell can be found elsewhere.\cite{Ref34,Ref35} The spectral weight on the constant-energy plane of 1.0 eV below $E_F$ are presented with different sizes of Brillouin zones sketched in Figs.~\ref{fig:fig2} (a)-(d). For the ($\sqrt{3}\times\sqrt{3}$) unit cell, a periodic pattern corresponding to a small Brillouin zone is obtained (Figs.~\ref{fig:fig2} (e) and (i)). Once the unit cell is chosen as the ($1\times1$) unit cell, shown in Fig.~\ref{fig:fig2} (b), the spectral weight shows a different periodicity due to a higher degree of freedom (Figs.~\ref{fig:fig2} (f) and (j)). This indicates that the translational symmetry breaking to ($1\times1$) silicene leads to shadow bands and distributes the weight unevenly. In Fig.~\ref{fig:fig2} (c), we further introduce an even smaller unit cell that contains a single Si atom. Then, the spectral weight is even more distributed (Figs.~\ref{fig:fig2} (g) and (k)) as compared to the previous two cases. For example, the intensity at the zone center is reduced and some spectral weight disappears almost completely. However, when choosing even half of the lattice constant of that in Fig.~\ref{fig:fig2} (c), the spectral weight is not altered, as shown in Figs.~\ref{fig:fig2} (h) and (l), although the overall spectral weight is reduced by a factor of four. This is required to satisfy the sum rule. Note that once the unfolded spectral weight is folded into the smallest Brillouin zone related to the original ($\sqrt{3}\times\sqrt{3}$) unit cell, the same spectral weight is \textit{exactly recovered}.

\begin{figure}[tbp]
\includegraphics[width=0.70\columnwidth,clip=true,angle=0]{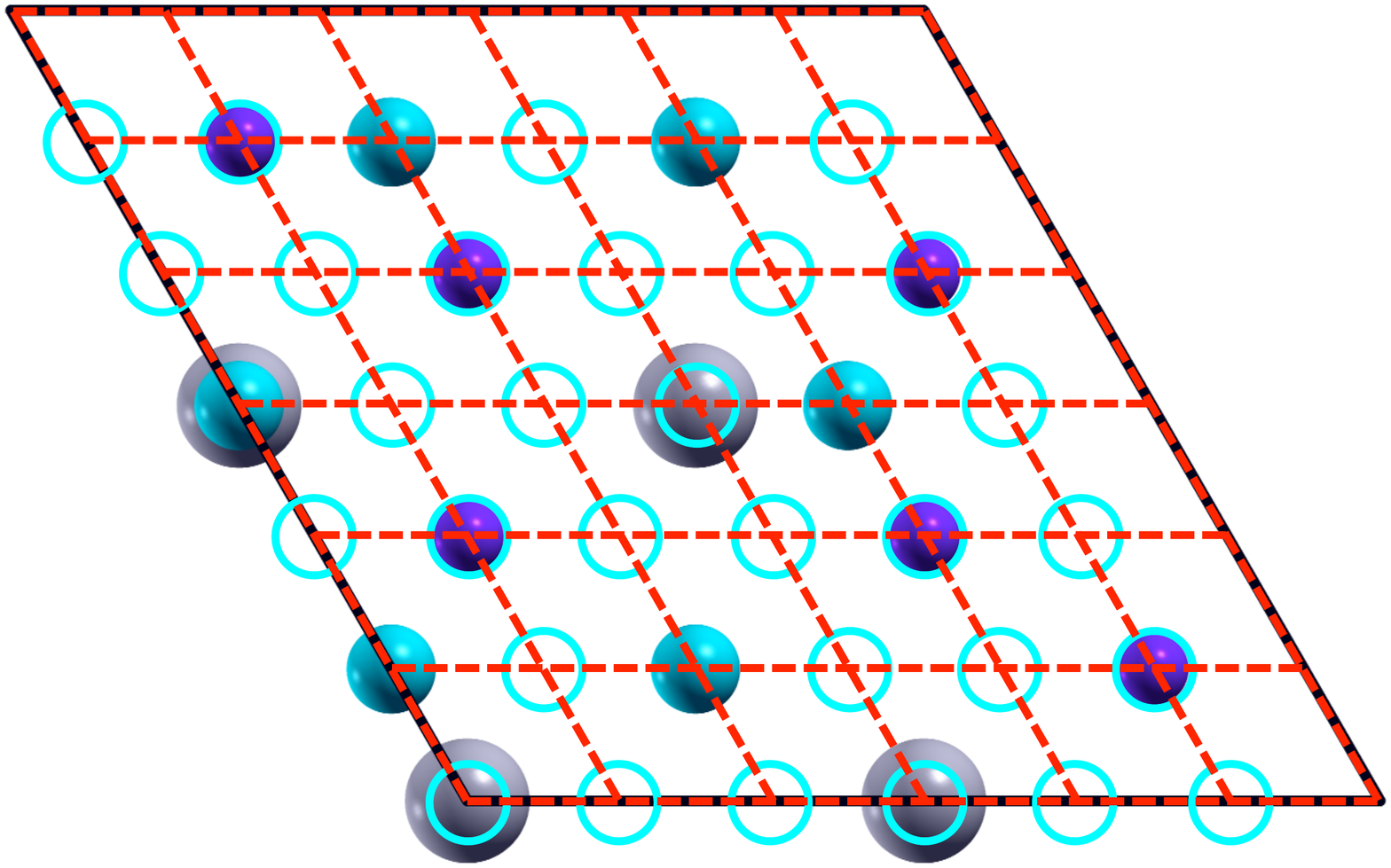}
\caption{\label{fig:fig3}
The ($6\times6$) supercell with respect to the conceptual ``one-Si-atom'' in-plane unit cell, which is presented by the dashed lines, reproduces the ($\sqrt{3}\times\sqrt{3}$)-reconstructed silicene on ZrB$_2$(0001). The open circles illustrate the ordered Si vacancies. The vacancies can also be applied for the Zr and B atoms. Note that the dashed unit cell contains only one atom per layer and is commensurate with both silicene and the ZrB$_2$ substrate.
}
\end{figure}

If the one-atom unit cell (Fig.~\ref{fig:fig2} (c)) would be a reasonably good choice to allow for the presentation of translational symmetry breaking in a large BZ, the commensurability for a complicated system becomes an important issue. For silicene on ZrB$_2$(0001), a good choice of the conceptual unit cell that contains only a single Si atom, a single Zr atom, and a single B atom for monolayer silicene and the topmost ZrB$_2$ layer of the substrate, respectively, is shown by dashed lines in Fig.~\ref{fig:fig3}, as this is the largest commensurate one-atom unit cell. In the following, this unit cell is referred to as the ``one-Si-atom unit cell''. In other words, the ($\sqrt{3}\times\sqrt{3}$)-reconstructed silicene can be reproduced upon removal of some ordered atoms, as vacancies, from the ($6\times6$) supercell. As expected, the unfolded spectral weight reflects the strength of the symmetry breaking due to the vacancies and out-of-plane deviations of the atomic positions.

\subsection{The calculated electronic band structure of silicene on ZrB$_2$(0001)}

Due to the epitaxial relationship to the Si(111) substrates, the experimentally observed in-plane lattice constant of ZrB$_2$(0001) thin films is somewhat larger (3.187 \AA) as compared to the experimentally observed (3.169 \AA) and calculated (3.174 \AA) \cite{Ref33} bulk values. Even with this larger lattice constant, free-standing silicene prefers an in-plane lattice constant that is longer than that allowed by the epitaxial conditions imposed by the ZrB$_2$(0001) surface.\cite{Ref5} Practically, as suggested by the observation of a regular arrangement of stripe domains in STM images\cite{Ref8} and by the relaxation of Si$_C$ atoms away from energetically unfavorable on-top positions as suggested by Si 2$p$ core-level spectra,\cite{Ref26} strain in silicene is released by structural relaxation accompanied by a small reduction of the areal density of Si atoms as compared to the system without stripe domains. While calculations that account for stripe domains require a huge supercell, a simpler way to simulate the reduction of the areal density is to explore larger in-plane lattice constants of the whole system. Additionally, the approximation for the exchange-correlation energy as used in the GGA is known to possibly overestimate the band width and could also give as much as up to 2\% error in the lattice constants.\cite{Ref37,Ref38,Ref39} In the following, we therefore compare the results obtained for $a$ = 6.348 \AA\ to those for the 2\%-longer lattice constant of $a$ = 6.480 \AA, where $a$ is the lattice constant of the ($2\times2$) unit cell of the ZrB$_2$(0001) reconstructed surface that corresponds to the ($\sqrt{3}\times\sqrt{3}$) unit cell of silicene.

\begin{figure}[tbp]
\includegraphics[width=0.98\columnwidth,clip=true,angle=0]{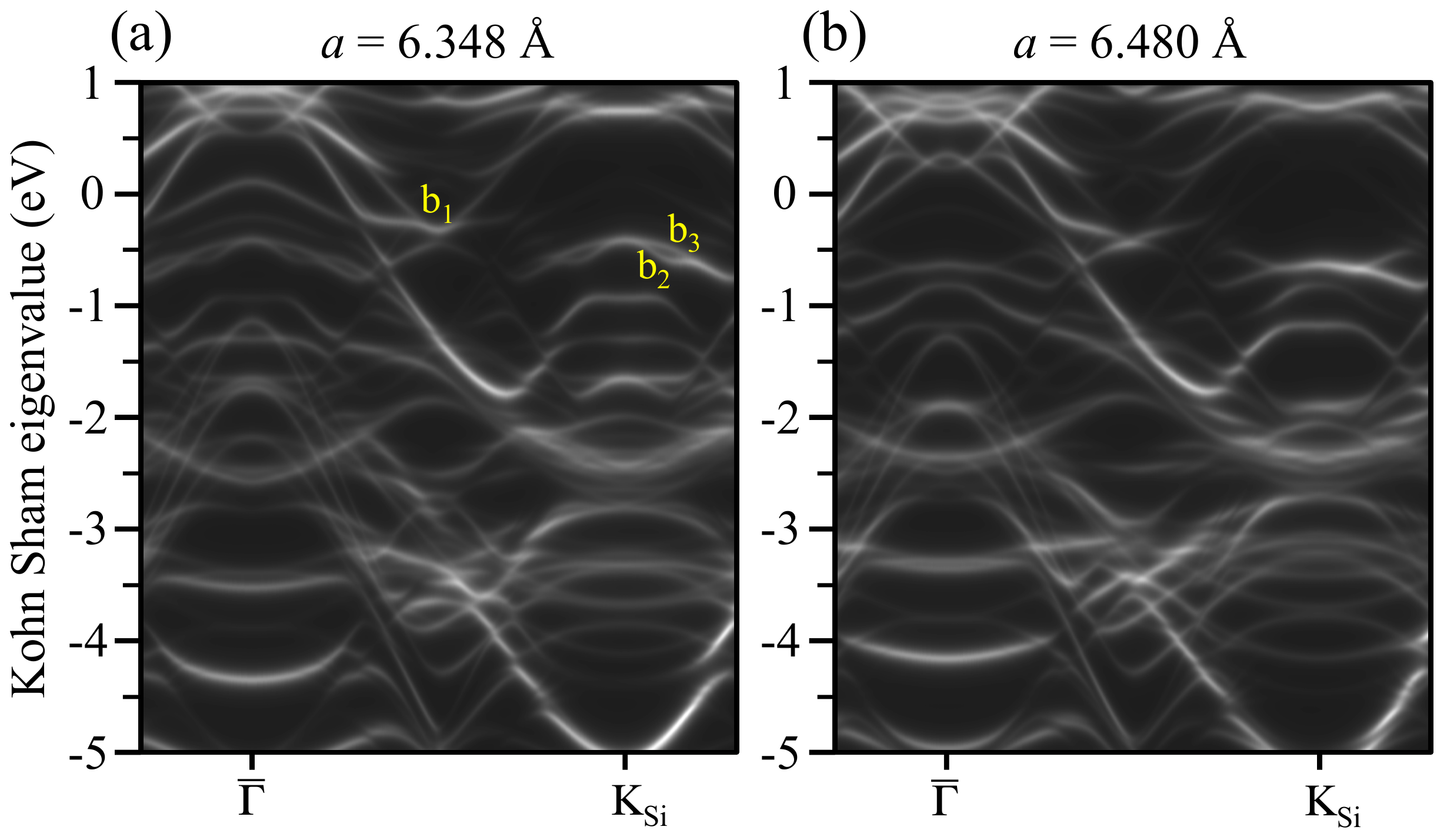}
\caption{\label{fig:fig4}
Electronic band structures of the buckled-like phase along the $\overline{\Gamma}$-$K_{Si}$ direction at two lattice constants expressed as the spectral weight of the electronic states: (a) $a$=6.348 \AA, and (b) $a$=6.480 \AA. Only the contribution from silicene and the terminating Zr layer are shown.
}
\end{figure}

\begin{figure}[tbp]
\includegraphics[width=0.98\columnwidth,clip=true,angle=0]{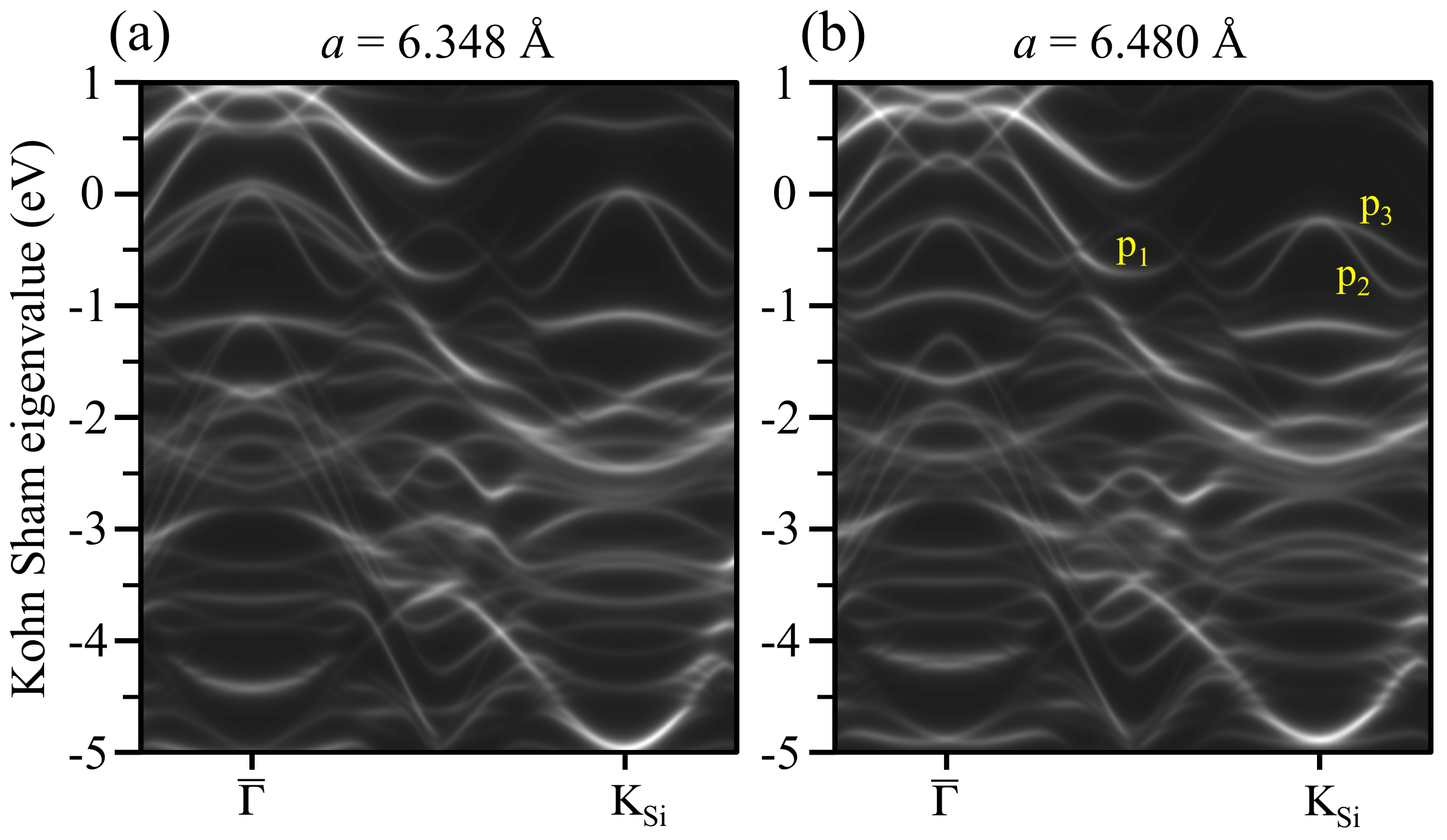}
\caption{\label{fig:fig5}
Electronic band structures of the planar-like phase along the $\overline{\Gamma}$-$K_{Si}$ direction at two lattice constants expressed as the spectral weight of the electronic states: (a) $a$=6.348 \AA, and (b) $a$=6.480 \AA. Only the contribution from silicene and the terminating Zr layer are shown.
}
\end{figure}

The electronic band structures represented in the BZ of the chosen commensurate conceptual one-Si-atom unit cell are shown in Figs.~\ref{fig:fig4} and~\ref{fig:fig5}, for the buckled-like and planar-like phases, respectively. It should be mentioned that in this representation, the $K$ point is at high values of the in-plane component of the electron momentum, $k_{\parallel}$, while in the Figures, we indicate the $K_{Si}$($1\times1$) point of non-reconstructed silicene. Here, in order to allow a comparison with the intensity of ARPES spectra, each band is weighted by its spectral weight as derived from the imaginary part of the one-particle Kohn-Sham Green's function.\cite{Ref34,Ref35} 

In this new representation, the differences between the band structures of buckled-like and planar-like epitaxial silicene phases are striking and substantial. Here, we focus on some of the most prominent and most conclusive features that are best recognized along the $\overline{\Gamma}$-$K_{Si}$ direction. In particular, while for both phases, the top of the two upward curved features in the vicinity of $E_F$ at the $K_{Si}$ point, shift down with increasing $a$, in the case of the buckled-like phase, they form a flatter feature for $a$ = 6.480 \AA\ but remain similar upward curved in the planar-like phase. The width of these bands, denoted $b_2$ and $b_3$ or $p_2$ and $p_3$, for the buckled-like and planar-like phases, respectively, is reduced with $a$ which reflects the expansion of the bond length at the longer lattice constant. Note that these features bear some resemblance to the predicted Dirac cone of $\pi$ bands of free-standing, non-reconstructed silicene and are therefore of particular interest. A parabolic band $p_1$ is only present in the planar-like phase where its energy and dispersion is almost unaffected by the choice of $a$. In contrast, the band $b_1$ of the buckled-like phase shows a v-like shape and the bottom weight is missing at $a$ = 6.480 \AA. Many bands at higher binding energy are clearly similar for both phases, and are thus certainly be related to the bulk of the diboride substrate.

\subsection{Comparison with ARPES spectra}

\begin{figure*}[tbp]
\includegraphics[width=1.75\columnwidth,clip=true,angle=0]{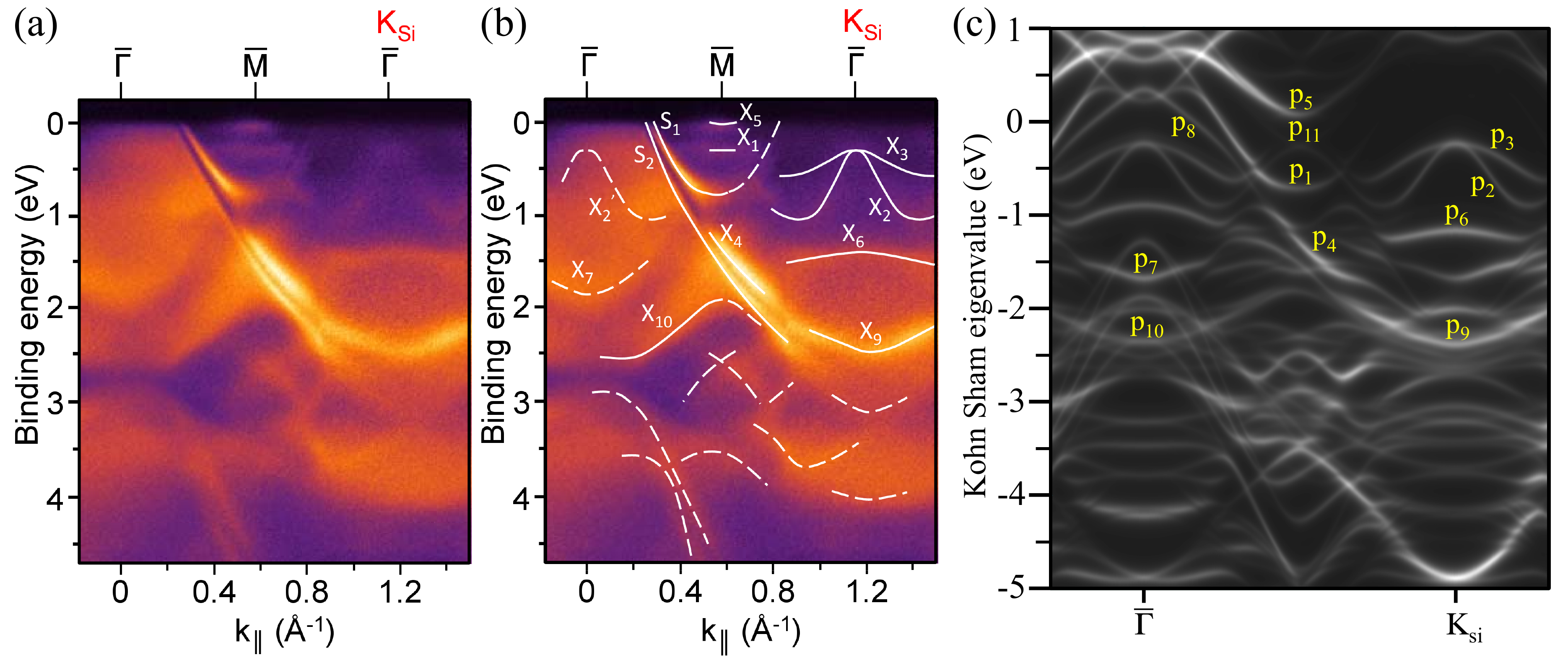}
\caption{\label{fig:fig6}
(a) ARPES spectra of epitaxial silicene on the ZrB$_2$(0001) surface along the $\overline{\Gamma}$-$K_{Si}$ direction. (b) ARPES spectra with guiding curves. (c) DFT band structure of planar-like phase at $a$ = 6.480 \AA.
}
\end{figure*}

ARPES spectra obtained along the $\overline{\Gamma}$-$K_{Si}$ direction and in a wide energy range, are shown in Figs.~\ref{fig:fig6} (a) and (b). The spectra are similar to those reported previously for the low-binding energy region.\cite{Ref8,Ref20} Accordingly, we use the same notation to mark the individual features. While $S_1$ and $S_2$ have been assigned to diboride surface states,\cite{Ref8} features $X_1$-$X_5$ are related to the presence of silicene.\cite{Ref8,Ref20} Feature $X_2$ is mirrored as $X_2^{\prime}$ at $\overline{\Gamma}$ thus following the symmetry of the ($\sqrt{3}\times\sqrt{3}$)-reconstructed silicene lattice. While $X_5$ has previously only been observed upon donation of electrons following the adsorption of potassium atoms,\cite{Ref20} interestingly, in the present spectra, possibly due to the higher count rate caused by the increased flux of the undulator photon source, it can be recognized in the pristine sample. The bottom of the associated otherwise unoccupied band is therefore pinned at $E_F$.

The comparison of the ARPES spectra with the band structure calculated for the planar-like phase at $a$ = 6.480 \AA\ shows an astonishing and excellent quantitative agreement in terms of the energies and dispersions of electronic states. In particular, $p_2$ and $p_3$ clearly find their counterpart in $X_2$ and $X_3$ while the unoccupied band $p_5$ almost touches $E_F$ exactly halfway in between the $\overline{\Gamma}$ and $K_{Si}$($1\times1$) points where feature $X_5$ is observed. The same good match is obtained between $p_{11}$, $p_4$, $p_6$, $p_7$, $p_1$, $p_8$, and $p_9$ and $X_1$, $X_4$, $X_6$, $X_7$, $S_1$, $S_2$ and $X_9$, respectively.

On the other hand, only partial agreement is found for the buckled-like phase, especially due to the lack of similarity between $b_1$ and $S_1$. The band widths of $b_2$ and $b_3$ are also too narrow to compare with the experimental observations. The present results are therefore strong evidence that the structural properties of silicene on the ZrB$_2$(0001) surface are very close to those of the planar-like phase that has been calculated to be the ground state stabilized by interactions between the Si honeycomb lattice and the metallic diboride surface.\cite{Ref5} In this context, it should again be emphasized that this could only be recognized now because of the implementation of two new methodological steps: (i) folding of bands is avoided by the choice of the one-Si-atom commensurate unit cell; and (ii) the exploration of larger in-plane lattice parameters.

The better agreement observed by enlarging $a$ may originate from the actual large lattice constant of the ZrB$_2$ thin films or from the possibly reduced surface density of the Si atoms (which is reduced in order to avoid epitaxial strain) but it may also originate from the DFT calculation itself. We used GGA as an approximation for the exchange correlation energy in first-principles calculations. While the exact form of the exchange-correlation energy is unknown, it has been discussed that the approximations could give rise to a 2\% error in the estimation of the lattice constant\cite{Ref38} and to a bandwidth renormalization factor of 3 as compared to experiments.\cite{Ref39} By enlarging the lattice constant, the Zr $d$ orbitals are effectively more localized. This in turn can give rise to a narrower band width as discussed before for approximations in the exchange-correlation energy.\cite{Ref37,Ref39} Note that this could also cause a too small band gap. But even if the deviations in the exchange-correlation energy are accounted for, the agreement is quite remarkable since due to the presence of stress-related stripe domains, the real atomistic structure of silicene on the ZrB$_2$(0001) surface is more complex than the one considered here.\cite{Ref8}

\subsection{The orbital character of the electronic states}

\begin{figure*}[tbp]
\includegraphics[width=1.70\columnwidth,clip=true,angle=0]{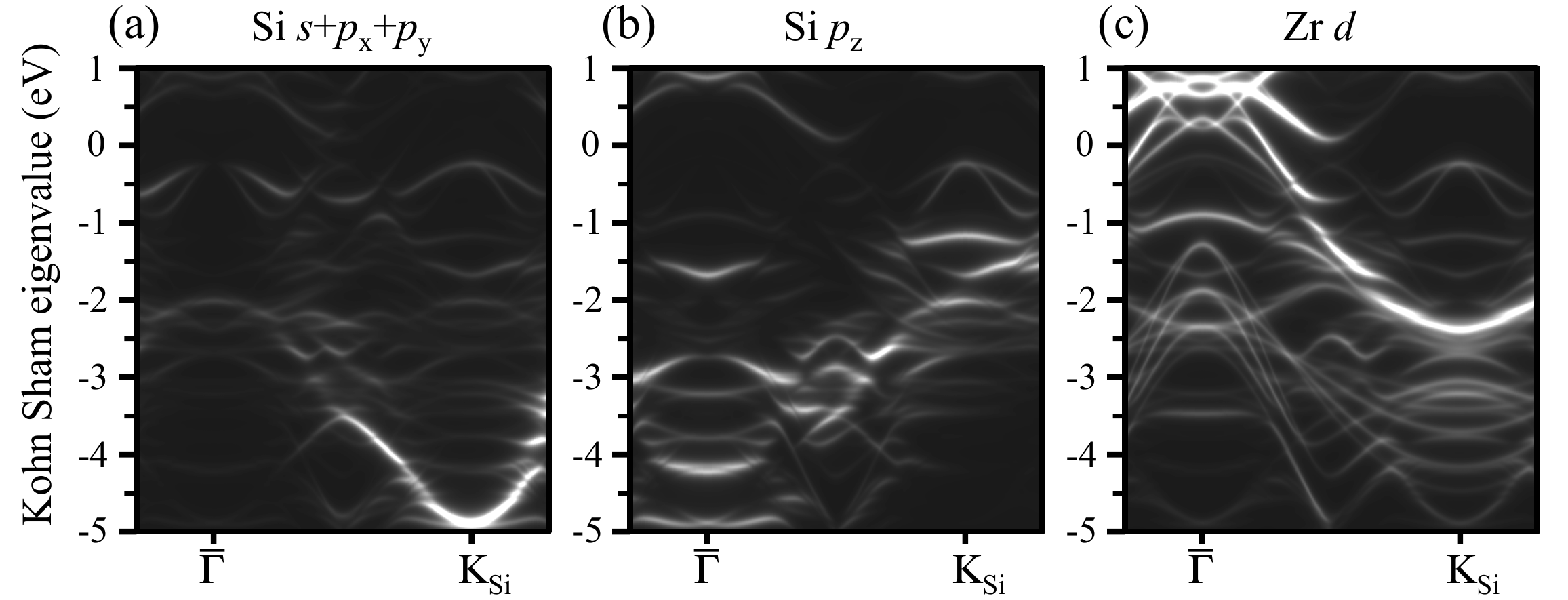}
\caption{\label{fig:fig7}
The (a) Si $s+p_x+p_y$, (b) Si $p_z$, and (c) Zr $d$ orbital character of the electronic bands of the planar-like phase at $a$=6.480 \AA\ along the $\overline{\Gamma}$-$K_{Si}$ direction. Zr $d$ denotes the contribution from the terminating layer that more strongly interacts with silicene.
}
\end{figure*}

Once settled that the band structure of DFT-proposed planar-like phase is consistent with the experimentally obtained ARPES data, the orbital character of the electronic states in this representation may be discussed. The individual contribution from Si $s+p_x+p_y$, Si $p_z$ and Zr $d$ orbitals are shown in Figs.~\ref{fig:fig7} (a), (b) and (c), respectively. All silicene-derived bands are hybridized to some extent with Zr $d$ electronic states, which is consistent with non-negligible interactions at the interface. In particular, while the upward curved bands $p_2$ and $p_3$ in the vicinity of the $K_{Si}$($1\times1$) point (that correspond to the spectral features $X_2$ and $X_3$ in the ARPES spectra) have major contributions from Si $p_z$ orbitals indeed, at the same time, they hybridize with the Si $s$, $p_x$, $p_y$ and the Zr $d$ orbitals. As already suggested in our previous work,\cite{Ref8} these states are therefore of partial $\pi$ character. Importantly, the $\pi$ character is robust against the choice of the larger Brillouin zone. Bands $p_6$ and $p_7$ ($X_6$ and $X_7$) also have strong contributions of $p_z$ orbitals and can be classified therefore as $\pi$ bands. On the other hand, bands $p_1$ and $p_8$ correspond to $S_1$ and $S_2$ in the ARPES spectra and have almost sole contributions from $d$ orbitals of the outermost Zr layer. This confirms that these states are diboride surface states indeed which survive upon the formation of the silicene layer, as has been concluded previously from experimental observations.\cite{Ref20} Band $p_5$ ($X_5$) is a hybrid state with contributions from Si $p_z$ and Zr $d$ orbitals. Note that upon K atom adsorption, electrons are donated to $X_5$ and not to $S_1$.\cite{Ref20} This is consistent with the partial $\pi$ character of $X_5$ and sole electron donation to silicene-derived states.

Our analysis shows that both bands $p_1$ and $p_5$ derive from combinations of the parabolic surface state of the pristine zirconium diboride surface\cite{Ref5} with either Si $p_z$ or Si $s+p_x+p_y$ orbitals, for $X_5$ and $S_1$, respectively. In other words, they are formed by rather non-local interactions between silicene and the substrate surface. On the other hand, if the Si$_C$ atoms are too close to Zr atoms, as they are in the buckled-like phase, local interactions destroy the diboride surface state.

Beside hybridization between silicene-derived and Zr-derived orbitals, we can recognize that there is also an apparent hybridization among Si $s$, $p_x$, $p_y$ and $p_z$ orbitals which reflects the intermediate $sp^2/sp^3$ hybridization of epitaxial silicene. Due to the complex hybridization, Si $p_z$ orbital contributions are spread out and occur up to binding energies of about 5 eV which is wider than the $\pi$ band width predicted for free-standing silicene of about 3-3.5 eV.\cite{Ref3,Ref4}

\section{Conclusions}

The combination of ARPES data obtained in a wide binding energy range, and the results of first-principles calculations allow for a yet unprecedented verification of the structural and electronic properties of epitaxial silicene on the ZrB$_2$(0001) surface. In order to obtain best agreement with the experimental data and to avoid folding of bands, in the calculations, the small and commensurate conceptual unit cell containing a single Si atom, a single Zr atom and a single B atom (as well as Si, Zr and B vacancies) is chosen. Additionally, the in-plane lattice constant has been slightly increased in order likely to account for a misestimate of the exchange-correlation energy in the GGA and to simulate the effect caused by the larger lattice constant of ZrB$_2$ thin films and the expected lower surface density of the Si atoms induced to avoid epitaxial strain.

Even if the actual stripe pattern observed in STM experiments\cite{Ref8} is not accounted for, the excellent agreement between the experimental and calculated band structures are strong evidence that the structural properties of silicene on the ZrB$_2$(0001) surface are very close to those of the planar-like phase. The results thus demonstrate that both calculations and experiment converge to this phase, which is stabilized by interactions between the Si honeycomb lattice and the metallic diboride surface.\cite{Ref5}

Finally, we have confirmed that all silicene-derived bands are hybridized to some extent with Zr $d$ electronic states, which is consistent with non-negligible interactions at the interface. But on the other hand, the zirconium diboride surface states survive upon the formation of silicene at the surface which indicates that these interactions are rather non-local. While the upward curved bands in the vicinity of the Fermi level at the $K_{Si}$($1\times1$) point are of partial $\pi$ character, they are hybridized with Si $s$, $p_x$, $p_y$ and Zr $d$ orbitals as well. This notable hybridization among Si $s$, $p_x$, $p_y$ and $p_z$ orbitals reflects the intermediate $sp^2/sp^3$ hybridization of epitaxial silicene.

Our new method of unfolding the band structures of the epitaxial layer and its substrate into a common conceptual unit cell enabled us to closely compare the electronic structures obtained theoretically and experimentally. The results therefore resolve the previously controversial issue related to the structural configuration of silicene on the ZrB$_2$(0001) surface and its band structure. The method is anticipated to be a powerful tool for the study of the electronic structures of epitaxial two-dimensional materials in general.

\section*{Acknowledgements}

We are grateful for experimental assistance from K. Mase (Institute of Materials Structure Science, KEK, Tsukuba, Japan), and for the use of the Cray XC30 machine at JAIST. The computational work was supported by the Strategic Programs for Innovative Research (SPIRE), MEXT, the Computational Materials Science Initiative (CMSI), and Materials Design through Computics: Complex Correlation and Non- Equilibrium Dynamics, A Grant in Aid for Scientific Research on Innovative Areas, MEXT, Japan. The experiments were performed under the approval of the Photon Factory Advisory Committee (Proposal No. 2012G610). YYT acknowledges financial support from the Funding Program for Next Generation World-Leading Researchers (GR046).
 
\bibliography{refs}
\end{document}